\documentclass[10pt]{iopart}
\usepackage{iopams}
\usepackage{amssymb}
\usepackage{graphicx}
\eqnobysec
\newcommand{\abs}[1]{\left\vert#1\right\vert}
\newcommand{\norm}[1]{\left\Vert#1\right\Vert}
\newcommand{\paren}[1]{\left(#1\right)}
\newcommand{\eps}{\epsilon}

\newcommand{\dz}{\partial_{z} }
\newcommand{\dx}{\partial_{x} }

\newcommand{\dzz}{\partial_{z}^{2}}
\newcommand{\zz}{{zz}}

\newcommand{\h}[1]{{H^{#1}}}

\newcommand{\dt}{\partial_{t}}
\newcommand{\f}{\phi}
\newcommand{\X}{\mathcal{X}}
\newcommand{\nn}{\nonumber}

\newcommand{\E}{\mathcal{E}}

\newcommand{\R}{\mathbb{R}}
\newcommand{\U}{\mathcal{U}}
\newcommand{\Low}{\mathcal{L}}
\newcommand{\qed}{\begin{flushright}{$\Box$}\end{flushright}}
\newtheorem{thm}{Theorem}[section]
\newtheorem{cor}[thm]{Corollary}
\newtheorem{lem}[thm]{Lemma}
\newtheorem{prop}[thm]{Proposition}
\newtheorem{defn}[thm]{Definition}

\begin{document}
\title[Degenerate Dispersive Equations]{Degenerate dispersive equations arising in the study of magma dynamics}
\author{G. Simpson$^1$, M. Spiegelman$^{1,2}$, M.I. Weinstein$^1$}
\address{$^1$ Department of Applied Physics and Applied Mathematics, Columbia University, New York, NY 10027,
USA}
\address{$^2$ Department of Earth and Environmental Science, Columbia University, New York, NY 10027
USA}
\eads{\mailto{grs2103@columbia.edu}, \mailto{mspieg@ldeo.columbia.edu}, \mailto{miw2103@columbia.edu}}

\begin{abstract}
An outstanding problem in Earth science is understanding the method of transport of magma in the Earth's mantle.  Two proposed methods for this transport are percolation through porous rock and flow up conduits.  Under reasonable assumptions and simplifications, both means of transport can be described by a class of degenerate nonlinear dispersive partial differential equations of the form:
\[
        \f_{t}+(\f^{n})_{z}-(\f^{n}(\f^{-m}\f_{t})_{z})_{z}=0,
\]
where $\f(z,0)>0$ and $\f(z,t)\to1$ as $z\to\pm\infty$.
Although we treat arbitrary $n$ and $m$, the exponents are physically expected to be between $2$ and $5$ and $0$ and $1$, respectively.

In the case of percolation, the magma moves via the buoyant ascent of a less dense phase, treated as a fluid,
through a denser, porous phase, treated as a matrix.  In contrast to  classical porous media problems where the matrix is fixed and the fluid is compressible, here the matrix is deformable, with a viscous constitutive relation, and the fluid is incompressible.  Moreover, the matrix is
modeled as a second, immiscible, compressible fluid to mimic the process of dilation of the pores.  Flow via a conduit is modeled as a viscously deformable pipe of magma, fed from below.

Analog and numerical experiments suggest that these equations behave akin to KdV and BBM; initial conditions evolve into a collection of solitary waves and dispersive radiation.  As $\phi\to 0$, the equations become degenerate.  A general local well-posedness theory is given for a physical class of data (roughly $H^1$)  via fixed point methods. The strategy requires positive lower bounds on $\f(z,t)$. The key to global existence is the persistence of these bounds for all time.  Furthermore, we construct a Lyapunov energy functional, which is locally convex about the  uniform porosity state, $\f\equiv1$, and  prove (global in time) nonlinear dynamic stability of the uniform state for {\it any} $m$ and $n$. For data which are large perturbations of the uniform state, we prove global in  time well-posedness for restricted ranges of $m$ and $n$. This includes, for example, the case $n=4,m=0$, where an appropriate uniform in time lower global bound on $\f$ can be proved using the conservation laws.  We compare the dynamics to that of other problems and discuss open questions concerning a larger range of exponents, for which we conjecture global existence.
\end{abstract}
\ams{74J30, 35A05}
\submitto{NL}
\maketitle
\section{Introduction and Overview}
An outstanding problem in Earth science is understanding the method of transport of magma in the Earth's mantle.  One proposed method is buoyant ascent of the melt through a viscously deformable porous medium.  Such a model may be visualized in \Fref{fig:mechanics} (a), where the magma flows through tubules along the grain boundaries of the solid rock.
\begin{figure}
        \begin{center}
\label{fig:mechanics}
\end{center}
 \caption{\bf{(a)} A single Olivine crystal, and the channels along which molten rock flows, \cite{Zhu:2003fk}.  \bf(b) Fluid flow up a viscously deformable pipe.}
\end{figure}
Equations for this process were derived independently in  \cite{McKenzie:1984ex} and \cite{Scott:1984sx, Scott:1986xs}.  In both derivations, the system is modeled using a volume averaged two phase flow, one for the melt and one for the matrix, at an extremely low Reynolds number.  This amounts to having two coupled conservation of mass equations to two force balance equations,
\begin{eqnarray*}
                \dt(\rho_f \f)+\nabla \cdot (\rho_f\f \vec{v}_f)=0\\
                \dt(\rho_s (1-\f))+\nabla \cdot (\rho_s(1-\f) \vec{v}_s)=0\\
                \nabla \cdot (\f \bold{\sigma_f} )= \rho_f \f \vec{g} +\vec{I}\\
                \nabla \cdot ((1-\f)\bold{\sigma_s}) = \rho_s (1-\f)\vec{g}-\vec{I}
\end{eqnarray*}
The variables are defined in \tref{table:parameters}.
\begin{table}[htdp]
        \caption{Viscously Deformable Media Variables}
        \begin{center}
                \begin{tabular}{lc}
                        \hline
                        Parameter & Symbol\\
                        \hline
                        Porosity & $\f$\\
                        Fluid density & $\rho_f$\\
                        Solid density & $\rho_s$\\
                        Fluid velocity & $\vec{v}_f$\\
                        Solid velocity & $\vec{v}_s$\\
                        Fluid rheology & $\sigma_f$\\
                        Solid rheology & $\sigma_s$\\
                        Gravitational vector & $\vec{g}$ \\
                        Interphase force & $\vec{I}$\\
                        \hline
                \end{tabular}
        \end{center}
        \label{table:parameters}
\end{table}%
The fluid rheology is that of an incompressible inviscid fluid while the solid rheology is treated as viscously compressible fluid.  We refer the reader to \cite{McKenzie:1984ex,Scott:1984sx,Scott:1986xs,Barcilon:1986vf, Nakayama:1992fk} for expositions on these relationships and their simplifications.

When reduced to 1-D, as in \cite{Scott:1984sx, Scott:1986xs, Barcilon:1986vf, Barcilon:1989qf,Spiegelman:1993pd, Spiegelman:1993ky}, this transport can be described by a class of degenerate nonlinear dispersive partial differential equations of the form:
\begin{equation}\label{eq:magmaequation}
        \paren{\f^m-\dz\f^n\dz} [\f^{-m}\f_t]=-(\f^n)_z
\end{equation}
This can be rewritten as the coupled system,
\begin{eqnarray}
        \label{eq:coupledsystema}
        \f_t = \f^m u \\
        \label{eq:coupledsystemb}
        \paren{\f^m-\dz\f^n\dz} u = -(\f^n)_z
\end{eqnarray}
The exponent $n$ corresponds to the power in the permeabily-porosity relationship
\[
        K\propto \f^n
\]
while the exponent $m$ corresponds to the power in the bulk viscosity-porosity relationship
\[
        \eta \propto \frac{1}{\f^m}
\]
In \cite{Scott:1984sx, Scott:1986xs}, the authors conclude that the parameter space for these exponents is $2\leq n \leq 5$ and $0\leq m \leq 1$.  In \cite{McKenzie:1984ex}, the bulk viscosity is taken to be constant, corresponding to $m=0$, and $2\leq n \leq 3$.  Several papers \cite{Barcilon:1986vf, Barcilon:1989qf, Spiegelman:1993ky} have focused on the case $n=3$ and $m=0$.

A second important mechanism for magma migration is that of flow through a viscously deformable pipe embedded in a viscous matrix, as takes place in the thermal plumes of the convecting mantle.  This is pictured in \Fref{fig:mechanics} (b).  The primitive equations, \cite{Olson:1986zr},
\begin{eqnarray*}
\dt A &=& -\dz Q\\
Q &=& \frac{A^2}{8 \pi \eta_L} \paren{\Delta \rho g + \eta_M \dz\paren{A^{-1}\dz Q}}
\end{eqnarray*}
whose variables are defined in \Tref{table:conduitparameters}, give rise to an equation of the form \eref{eq:magmaequation}, with $n=2$ and $m=1$, replacing $\phi$ with $A$. 
\begin{table}[htdp]
        \caption{Conduit Flow Variables}
        \begin{center}
                \begin{tabular}{lc}
                        \hline
                        Parameter & Symbol\\
                        \hline
                        Areal region of pipe & $A$\\
                        Flux up pipe & $Q$\\
                        Fluid viscosity & $\eta_L$\\
                        Matrix viscosity & $\eta_M$\\
                        Fluid-Matrix density difference & $\Delta \rho$\\
                        Gravitational constant & $g$ \\
                        \hline
                \end{tabular}
        \end{center}
        \label{table:conduitparameters}
\end{table}%
Analog laboratory models for this mechanism were studied in \cite{Olson:1986zr, Whitehead:1988fk}, verifying the theory by demonstrating the appearance and interaction solitary waves, with measurements consistent with the equations.  Throughout this paper, we will use $\f$ as the dependent variable.

Upon inspection, it is not clear what the behavior of \eref{eq:magmaequation} is.  It has both a nonlinear forcing term and a nonlinear dispersive term, in addition to being degenerate and nonlocal. The degeneracy is evident in  \eref{eq:coupledsystemb}, where the inevitability and elliptiticity of the operator break down as $\f\to0$.  {\it This mathematical challenge is at the heart  modeling the  underlying physical system; can the melt disaggregate in the case of porous flow, or can the conduits split in the case of pipe flow? Are there upper and lower bounds for physical variables such as porosity?}

Previous studies, \cite{Barcilon:1986vf, Barcilon:1989qf, Takahashi:1990tr, Takahashi:1992lp, Nakayama:1991ud, Nakayama:1994cq, Nakayama:1995hd, Nakayama:1999nb}, of the equations did not successfully address the issue of well-posedness and the related issue of a lower bound on $\f$, the (scaled) melt fraction, although \cite{Takahashi:1990tr} did point out several reasons why solutions for which $\f$ went to zero would be non-physical.  

Degenerate nonlinear {\it parabolic} partial differential equations are an important class of equations arising in porous media flow (with a fixed matrix) \cite{Aronson:1986kx} and in geometric flows, {\it e.g.} motion of a surface by its local curvature, \cite{Gage:1986fk}. However, the study of degenerate nonlinear {\it dispersive} equations appears to be wide open.  Our equations bear some resemblance to those considered in \cite{Rosenau:1993xh}, known to admit compactly supported solitary waves. 
We can see the dispersive character of the equation by linearizing about some background porosity $\f_0$.  The linearized equation
\[
        \f_t -n \f_0^{n-1} - \f_0^{n-m} \f_{zzt} =0
\]
has the dispersion relation
\[
        \omega(k) = \frac{n \f_0^{n-1} k}{1+\f_0^{n-m} k^2}
\]
with group velocity
\[
        \omega'(k) = n \frac{\f_0^{n-1} - \f_0^{2n-m-1}k^2}{(1+\f_0^{n-m} k^2)^2}
\]
When $\f_0=1$, which will be the background value of $\f$ in our analysis, we see the similarity between this equation and the Regularized Long Wave (RLW) equation, also known as the  Benjamin-Bona-Mahoney (BBM) equation
\[
        \dt u+ \dx u + u \dx u -\dx^2\dt u=0
\]
which has a linearized (about zero) dispersion relation
\[
        \omega(k) = \frac{k}{1+k^2}
\]
with group velocity
\[
        \omega'(k) = \frac{1-k^2}{(1+k^2)^2}
\]
The BBM equation \cite{Peregrine:1966vn, Benjamin:1972ms}, was proposed as an alternative to the Korteweg - de Vries (KdV) equation and is asymptotically equivalent to KdV in the regime of small amplitude long waves. Like KdV, BBM admits stable solitary waves, although it is not completely integrable.  The magma equations also have solitary waves, which play a central role in the general dynamics; see, for example,  \cite{Scott:1984sx,Olson:1986zr,Barcilon:1986vf,Scott:1986xs,Takahashi:1992lp,Barcilon:1989qf,Takahashi:1990tr,Nakayama:1991ud,Spiegelman:1993pd,Spiegelman:1993ky}. In this paper we focus on well-posedness questions and stability of the uniform state. The question of stability of solitary waves is currently under study.

The main results of this paper are the following:
\begin{itemize}
\item[(1)] A local well-posendess theory is established for arbitrary initial conditions, $\f(z,0)$, in a physically natural set of functions, ${\cal X}$,  of finite "energy", ${\cal E}$. The time-interval of existence, and therefore global  well-posedness, is shown to be controlled by $\|\f\|_{H^1}+\min_{z\in \R}\f(z,t)$. See Theorem \ref{thm:hklocalexistence}.
\item[(2)] We obtain ranges of material exponents, $(m,n)$, for which global well-posedness for arbitrary data in ${\cal X}$ holds. See Corollaries \ref{cor:smallh1globalexistence} and \ref{cor:largeh1globalexistence}, along with \fref{fig:parameterplot}. This range of parameters includes a set of obviously degenerate cases (such as $m=0$ and $n=4$), as noted in Corollary \ref{cor:degeneracy} and discussed in \Sref{sec:discussion}.  The important case of McKenzie's equation, $m=0$ and $n=3$ remains open. We conjecture global well-posedness.
\item[(3)] For {\it any} values of $m$ and $n$, we prove that the uniform porosity state, $\f(z,t)\equiv1$ is shown to be nonlinearly stable in Corollary \ref{cor:nonlinearstability}. The key step is to show that an appropriately constructed functional, defined on ${\cal X}$, is locally convex about the uniform state.  This strategy has been used in many Hamiltonian and nonlinear wave problems, see \cite{Arnold:1965kx, Benjamin:1972ol, Holm:1985fk, Weinstein:1986uq}.
\end{itemize}

\begin{figure}
        \begin{center}
                \caption{The trapezoid shows the permeabily-viscosity pairs of exponents, $(n,m)$, for which we have global existence for data of arbitrarily large $\h{1}$ norm. The line $n+3m=4$ bounds the region for       which we can ensure a lower bound, while the line $n+3m=6$ bounds the region            for which we can ensure an upper bound.  The line $n+m=3$ is a technical        constraint.  For all other $(n,m)$ pairs, we have global existence for data near the uniform state of $1$.}
                \label{fig:parameterplot}
        \end{center}
\end{figure}

In a forthcoming paper, we shall treat the question of stability of small amplitude solitary waves \cite{Simpson:uq}.

We proceed as follows.  In \sref{sec:localexistence} we exhibit local well-posedness in $H^k(\R)$ spaces for $k=1,2,\dots$  In \sref{sec:conservation} we discuss certain conserved quantities associated with the equations.  \Sref{sec:conservation}  addresses certain properties of a conserved functional and the nonlinear stability of the uniform background state.  In \sref{sec:smalldata} we prove global well-poesdness for data near the uniform background state, and in \sref{sec:largedata} we show global well-posedness for large data for a certain subset of the equations.  In \sref{sec:discussion} we make comparisons with other equations, conjecturing how the equation regularizes itself through the time dependent length scale of the problem.  We then summarize and discuss future work and open problems.

\bigskip

Throughout this paper, we shall use the following notation.
\begin{enumerate}
\item $H^k$ denotes the Sobolev space $W^{k,2}(\R)$ of functions with $L^2$ weak derivatives up to order $k$.  $k$ will always denote an integer, $k\geq1$.
\item $C^{k,\alpha}$ denotes the space functions with derivatives up to order $k$ are  H\"older continuous with exponent $\alpha$
\item All integrals are taken over $\R$ in the spatial variable or on some finite interval $[0,t]$ in the temporal variable.
\item Generic constants are denoted by $C,C',C''\ldots$ or $D,D',\ldots$
\item The dependence of a solution $\f$ of \eref{eq:magmaequation} on its data be may be explicitly expressed as
\[
	\f(z,t;\f_0)
\]
where $\f_0$ is the initial condition.
\end{enumerate}
 
\section{Local in time well-posedness in $\h{k}$,\ $k\ge1$}
\label{sec:localexistence}

In this section we consider the well-posedness of the initial value problem for \eref{eq:magmaequation}, existence, uniqueness, and continuous dependence with respect to the data.  Formally, solving \eref{eq:coupledsystemb} for $u$, we substitute this solution into \eref{eq:coupledsystema} and integrate with respect to $t$.  Applying an appropriate initial condition for $\f(z,t)$, we obtain
 \begin{eqnarray}
 \Lambda[\phi] &=& \phi_{0}+\int_{0}^{t} \f^m L_{\phi}^{-1}[-(\phi^{n})_{z}]ds,\label{Lambda-def}\\
L_{\phi}u&=&-(\phi^{n} u_{z})_{z}+\f^m u\label{Lphi-def}
\end{eqnarray}

The mapping $\Lambda$ is highly nonlinear, in no small part due to the  dependence of 
the nonlocal operator, $L_\f$, on $\f$.
Our strategy is to construct solutions of the initial value problem for  (\ref{eq:coupledsystema}-\ref{eq:coupledsystemb}) seeking a fixed point of the mapping $\f\mapsto\Lambda[\f]$ in an appropriate metric space of functions.  In particular, we shall apply the contraction mapping principle \cite{Rudin:1976ww} to $\Lambda$, restricted to an appropriately chosen closed subset  of  functions, $\X_{R,\eps,T}^{k}$, for which
 $\f(z,t)-1\in H^1$.  The set $\X_{R,\eps,T}^{k}$ is constructed so that
 iteration of $\Lambda$ on it preserves ellipticity and invertibility properties of $L_\f$.

\begin{defn}
\begin{enumerate}
\item\ Given  $1\ge \eps >0 $, $R,T>0$, $k\ge 1$, we define
\begin{eqnarray}  
\X_{R,\eps,T}^{k}& =&  \Big \{ \phi\ :\ \phi-1\in C^{1}([0,T],H^{s}(\R)) :\nn\\
& &\sup_{t<T}\norm{\phi(\cdot,t)-1}_\h{k}\leq R,\;  \sup_{t<T}\norm{\phi(\cdot,t)^{-1}}_\infty \leq \eps^{-1} \Big\}
\label{eq:space}
\end{eqnarray}
The sets $\X_{R,\eps,T}^{k}$ obviously include the constant solution, $\phi=1$, so they are nonempty and closed sets.
\item A metric on $\X_{R,\eps,T}^{k}$ is defined by
\begin{equation}
                [\phi-\psi]_k(T)=\sup_{t<T} \norm{\phi(\cdot,t)-\psi(\cdot,t)}_{H^{k}}
\label{eq:metric}
\end{equation}
\end{enumerate}
\end{defn}
We note that $\X$ is a metric space, not a function space; it is centered about the constant solution $1$.

\begin{defn}
We call $\phi$ a local solution of \eref{eq:magmaequation} with data $\f_0$ on the time interval $[0, T]$, if $\f>0$, $\f-1\in H^k$ and $\f = \Lambda[\f]$.
\label{def:solution}
\end{defn}

\begin{thm} (Local well-posedness) 
        \label{thm:hkfixedpoint}
        Let $1\geq\eps>0$, $R>0$, and $k \ge 1$ . Then, for $T>0$ sufficiently small, $\Lambda$ is a contraction on $\X_{R,\eps,T}^{k}$ with respect to the metric \eref{eq:metric} and hence admits a unique fixed point.
Thus, the initial value problem for (\ref{eq:coupledsystema}-\ref{eq:coupledsystemb}) has a unique local solution.  Moreover the solution depends continuously on the initial data.
\end{thm}

\noindent{\bf Remarks on notation:}\ Generic constants, $C$, may depend on $R,\eps,T$ and $k$, along with the two parameters in the equation $n$ and $m$.  While the dependence on $n$ and $m$ will suppressed in our proofs, the reader should be aware of it, and the change in the behavior of $C$ as $\eps\searrow 0$ and $R\nearrow \infty$ as $n$ and $m$ vary.  In the cases of geophysical interest, where $n>2$ and $0<m<1$, constants may behave like:
\[
	C\rightarrow\infty, \ {\rm as}\ \eps\searrow 0\ {\rm and}\ R\nearrow\infty
\]
Thus, \emph{ensuring that $\f$ stays away from zero is critical to our problem as it evolves in time.}
{\bf In the following, unless otherwise specified, $k\ge 1$, $R>0$ and $1\geq\eps>0$ are arbitrary.}
\noindent We now embark on setting up the proof of Theorem \ref{thm:hkfixedpoint}, which requires several technical steps. 

\begin{prop}
        \label{prop:bounds}
       Let $R,\eps,T>0$ and $k\ge 1$.  For all $\f \in \X_{R,\eps,T}^k$ 
        \begin{enumerate}
                \item $\f \in C^{k-1,1/2}(\R)$
                \item $\f\geq \eps$ 
                \item There exist $M_j<\infty$ such that $\norm{\partial_z^j\f}_\infty\leq M_j$ for $j\leq k-1$.  The $M_j$ depend on $R,\eps$ and $k$.
        \end{enumerate}
\end{prop}
Proof:  This follows from Sobolev inequalities \cite{Evans:2002ah} and the definition of $\X_{R,\eps,T}^k$.\qed
Also of use is that the Sobolev spaces $H^k(\R)$ form an algebra and admit the following inequality {\bf give citation}
\begin{lem}
        \label{lem:algebra}
       Assume $k>1/2$.  If $f,g \in H^k(\R)$, then
\begin{equation}
                \norm{f g}_{H^k} \leq C_k \norm{f}_{H^k} \norm{g}_{H^k}
\label{eq:sob-product}
\end{equation}
\end{lem}
The nonlinearity in the magma equations appears as a power of the dependent variable.  It will be useful to note the following property of these functions. 

\begin{prop}
        \label{prop:powerfunctions}
        Let $1\ge \eps>0$, $R,T>0$ and $k\ge 1$.  For all $p \in\R$ there exists a Lipschitz constant $C=C(R,\eps,k,p)$, such that for $\f,\psi\in \X_{R,\eps,T}^k$,
        \[
                \norm{\f(\cdot,t_1)^p-\psi(\cdot,t_2)^p}_{H^k} \leq C \norm{\f(\cdot,t_1)-\psi(\cdot,t_2)}_{H^k}
        \]
        for $t_1,t_2\leq T$.
\end{prop}
Proof:  This follows from the  bounds of Proposition \ref{prop:bounds},
  Lemma \ref{lem:algebra}  and that property that $x\mapsto x^p$ is $C^\infty$ for $x$ bounded away from zero.\qed
\begin{prop}
        \label{prop:invertibility}
        \begin{enumerate}
  \item \label{prop:invertibility:part1}    Let $R,\eps,T>0$ and $k\ge 1 $.  Let $\f\in \X_{R,\eps,T}^{k}$ and $f\in L^2$. Then,  $L_{\f(\cdot,t)}u=f$ has a unique solution $u \in H^{1}(\R)$ for $t\in[0,T]$.  Moreover, there exists a constant $C=C(R,\eps)$, such that
\begin{equation}
                \norm{u(\cdot,t)}_{H^{1}} \leq C \norm{f}_{L^2}
\label{eq:H1bound}
\end{equation}
\item     Furthermore, if $\f\in \X_{R,\eps,T}^{k+2}$ and $f\in H^k$, then $u(\cdot,t) H^{k+2}\in$, then there exists a constant $C=C(R,\eps,k)$ such that 
\begin{equation}
                \norm{u}_{H^{k+2}}\leq C \norm{f}_{H^{k}}
 \label{Hkp2bound}
 \end{equation}
        \end{enumerate}
\end{prop}
Proof:\  See \cite{Evans:2002ah} and \cite{John:1982vd} for details.  We  outline the arguments.
\begin{enumerate}
\item Since $\f\in \X_{R,\eps,T}^{k}$,  $L_\f$ is self-adjoint and positive definite. The bilinear form $\langle U,V\rangle_\f\ =\ 
 \int \f^n UV dz\ +\int \f^m \dz U\dz V\ dz$  defines  inner product on $H^1$. By the  Riesz Representation Theorem and there exists a unique $H^1$ solution, $u(\cdot,t)$, which satisfies  $\langle V ,u(\cdot,t)\rangle_\f\ =\ \int V f\ dz$ for all $V\in H^1$ .  Clearly, 
  (\ref{eq:H1bound}) holds.     
  \item Higher regularity may be deduced by first proving it for the case $k=0$ and then applying induction.  The $k=0$ case is proved by noting that the regularity of $\f$ implies the coefficients will be $C^{k+1}$ and studying the limit of difference quotients to  estimate higher derivatives. Alternatively, since the problem is in one spatial dimension, we could compute  and estimate derivatives of $u$ explicitly using variation of parameters representation of the $L_\f^{-1}$.
  \end{enumerate}
  \qed
\begin{prop}
        \label{prop:uhkbound}
       Let $1\ge \eps>0$, $R,T>0$ and $k\ge 1 \ge 1$.  There exists a constant $C=C(R,\eps,k)$ such that for all $\phi \in \X_{R,\eps,T}^k$, if $u$ solves $L_{\phi}u=-(\f^n)_z$, then
        \[
                \norm{u}_{H^k}\leq C
        \]
\end{prop}
Proof: The result follows by applying Proposition \ref{prop:invertibility} to $f= -(\f^n)_z\in \h{k-1}$.\qed
\begin{prop}
\label{prop:hkmap}
\begin{enumerate}
\item  Let $1\geq\eps>0$, $R>0$ and $k\ge 1$.  Assume $\norm{\phi_{0}-1}_\h{k}<R$ and $\f_0\geq 2\eps$.
        Then there exists $T_1=T_1(R,\eps,k,\f_0)>0$, such that for  $T<T_1$ , 
        \begin{equation}
        \f\in\X_{R,\eps,T}^{k}\ \ \longrightarrow\ \ \  \Lambda[\f]\in\X_{R,\eps,T}^{k}
\end{equation} \label{prop:Lambdastable}

 \item Moreover, there exists $0<T_2\le T_1$, $T_2=T_2(R,\eps,k,\f_0)>0$ such that for $T<T_2$ $\Lambda$ is a contraction on $\X_{R,\eps,T}^{k}$, {\it i.e.}, there exists an $\alpha=\alpha(R,\eps,k,\f_0)<1$ such that for all $\f,\psi \in \X_{R,\eps,T}^{k}$,
 \begin{equation}
 \label{eq:contraction}
 [\Lambda[\f]-\Lambda[\psi]]_k(T)\leq \alpha[\f-\psi]_k(T)
\end{equation}
 \label{prop:contract}
  \end{enumerate}
  \end{prop}
We first prove Proposition \ref{prop:hkmap} \eref{prop:Lambdastable}.  We must thus find a $T_1$ such that choosing $T<T_1$, for all $t \leq T$
\begin{equation}
	\norm{\Lambda[\f](\cdot,t)-1}_\h{k}\leq R
\label{eq:mapconditiona}
\end{equation}
\begin{equation}
\norm{\frac{1}{\Lambda[\f](\cdot,t)}}_\infty\leq \frac{1}{\eps}
\label{eq:mapconditionb}
\end{equation}
and
\begin{equation}
\Lambda[\f](\cdot,t)-1 \ge 1 C^1([0,T_1);\h{k})
\label{eq:mapconditionc}
\end{equation}

To establish \eref{eq:mapconditiona}, let  $u=L_{\phi}^{-1}[-(\phi^{n})_{z}]$. 
Then
\[
        \norm{\Lambda \phi-1}_{H^{k}}\leq \norm{\phi_{0}-1}_{H^{k}}+\int_{0}^{t}\norm{\f^m(\cdot,s)u(\cdot,s)}_{H^k}ds 
\]
Applying Lemma \ref{lem:algebra}, Proposition \ref{prop:powerfunctions}  and Proposition \ref{prop:invertibility} to the integrand  we get
\begin{equation}
\fl\norm{\f^m u}_{H^k} \leq C\norm{\f^m-1}_{H^k} \norm{u}_{H^{k}} + \norm{u}_{H^k}\leq C''(CC'\norm{\f-1}_{H^1} +1)\leq \tilde{C}
\label{eq:hkintegrandbound}
\end{equation}
Hence,
\[
        \norm{\Lambda \phi-1}_{H^{1}}\leq \norm{\phi_{0}-1}_{H^{1}} + \tilde{C}T
\]
Choosing $\norm{\phi_{0}-1}_\h{k}<R$, $T$ may be chosen sufficiently small such that $\norm{\Lambda \phi
-1}_{H^{k}}\leq R$. This shall be our first candidate for $T_1$, which we denote as $t_1$.

To prove \eref{eq:mapconditionb}, consider the difference between
$\phi_{0}$ and $\Lambda \phi$,
\[
\fl\norm{\Lambda\phi(\cdot,t)-\phi_{0}}_{\infty}\leq
\sqrt{2}\norm{\Lambda\phi(\cdot,t)-\phi_{0}}_{H^{k}}\leq\sqrt{2}\int_{0}^{t}\norm{\f^m(\cdot,s) u(\cdot,s)}_{H^{1}}ds\leq \sqrt{2} \tilde{C} T
\]
where $\tilde{C}$ is the constant in \eref{eq:hkintegrandbound}.  
\[
\abs{\Lambda[\f]}\geq \abs{\f_0}-\abs{\Lambda[\f]-\f_0}\geq \abs{\f_0}-\norm{\Lambda[\f]-\f_0}_\infty\geq 2\eps -\sqrt{2}\tilde{C}T
\]
Taking $T$ sufficiently small, this will be bounded below by $\eps$.  Hence, 
\[
        \Lambda [\phi] \geq \eps \qquad \mbox{or} \qquad         \norm{\frac{1}{\Lambda[\phi]}}_{\infty}\leq \frac{1}{\eps}\]

\eref{eq:mapconditionc} is more difficult to establish, requiring the following two Lipschitz-type estimates, which will also be needed in the proof of Proposition \ref{prop:hkmap} \eref{prop:contract}. This additional work is necessary because of the appearance of nonlinearity in the operator itself in contrast to more common nonlinear nonlocal dispersive equations.  We thus requires bounds on the "closeness" of the operators $L_\f$ for different $\f$'s, which we establish in the following propositions.

\begin{prop}
        \label{prop:hkoperatordifference}
        Assume $1\geq\eps>0$, $T,R>0$ and $k \ge 1$.  Let $\f$, $\psi$ be in $\X_{R,\eps,T}^k$.  Then for any two values $t_1,t_2<T$ at which $\f$ and $\psi$ are evaluated, there exists a constant $C=C(R,\eps,k)$ such that
        \[
                \norm{L_\phi^{-1}[-(\f^n)_z](\cdot,t_1)-L_\psi^{-1}[-(\psi^n)_z](\cdot,t_2)}_\h{k}\leq C\norm{\phi(\cdot,t_1)-\psi(\cdot,t_2)}_\h{k}
        \]
\end{prop}
Proof: Let $u$ and $v$ satisfy $L_{\f(\cdot,t_1)} u = -(\f(\cdot,t_1)^n)_z$ and $L_{\psi(\cdot,t_2)} v = -(\psi(\cdot,t_2)^n)_z$.  Then
\[
        L_\phi[u]-L_\psi[v] = -(\phi^{n}u_{z})_{z} + (\psi^{n}v_{z})_{z}+\phi^m u-\psi^m v = (\psi^n-\phi^n)_z
\]
Let $\gamma$ be in $\h{1}(\R)$.  Multiplying by $\gamma$ and integrating by parts,
\[
        \int (\phi^n u_z -\psi^n v_z) \gamma_z +(\phi^m u -\psi^m v)\gamma = \int (\psi^n - \phi^n) \gamma
\]
After adding and subtracting and  rearranging terms,
\begin{eqnarray}
        \label{eq:operatordifference}
       \int \phi^n(u-v)_z \gamma_z +\phi^m (u-v)\gamma =& \int& (\psi^n - \phi^n) \gamma\nonumber +(\psi^n-\phi^n)v_z\gamma_z\\
        & &+(\psi^m-\phi^m)v\gamma
\end{eqnarray}
For $k=1$, if we substitute $u-v$ for $\gamma$, we get the estimate
\begin{eqnarray*}
\int \phi^n((u-v)_z)^2 + \phi^m (u-v)^2 &\leq&\norm{\psi^n-\phi^n}_{L^2}\norm{u-v}_{L^2}\\
& &+\norm{\psi^n-\phi^n}_{L^\infty}\norm{v_z}_{L^2}\norm{(u-v)_z}_{L^2}\\
& &+\norm{\psi^n-\phi^n}_{L^\infty}\norm{v}_{L^2}\norm{u-v}_{L^2}
\end{eqnarray*}
which, after some further manipulation becomes,
\[
        \norm{u-v}_{H^1}\leq C \norm{\phi-\psi}_{H^1}
\]

For greater regularity, with $k\geq 2$, we return to \eref{eq:operatordifference}, and integrate by parts
\[
\fl\int \phi^n(u-v)_z \gamma_z +\phi^m (u-v)\gamma = \int \paren{-((\psi^n-\phi^n)v_z)_z+ \psi^n - \phi^n(\psi^m-\phi^m)v}\gamma
\]
This expresses $w=u-v$ as a solution to the elliptic problem $L_\phi w = f$ with
\[
f=-((\psi^n-\phi^n)v_z)_z+ \psi^n - \phi^n(\psi^m-\phi^m)v
\]
Indeed, for $k\geq 2$, we know $w$ is at least a weak solution, as $f\in L^2$, and the coefficients in the operator are bounded.  We may thus apply Proposition \ref{prop:invertibility} to get
\[
        \norm{u-v}_{H^{k}}\leq C \norm{f}_{H^{k-2}}\leq C \norm{\phi-\psi}_{H^{k}}
\]
\qed
We also need to be able to consider normed differences of $\f^mL_\phi^{-1}[-(\f^n)_z]$ for which we make the following estimate,
\begin{prop}
        \label{prop:hkratedifference}
        Assume $1\geq\eps>0$, $T,R>0$ and $k\ge 1$.  Given $\f$ $\psi$ in $\X_{R,\eps,T}^1$, there exists a constant $C=C(R,\eps,k)$ such that for any $t_1,t_2<T$
        \[
                \fl\norm{\f(\cdot,t_1)^mL_{\phi(\cdot,t_1)}^{-1}[-(\f(\cdot,t_1)^n)_z]-\psi(\cdot,t_2)^mL_{\psi(\cdot,t_2)}^{-1}[-(\psi(\cdot,t_2)^n)_z]}_{H^k}\leq C \norm{\phi(\cdot,t_2)-\psi(\cdot,t_2)}_{H^k}
        \]
\end{prop}
Proof: Let $u$ and $v$ satisfy $L_\phi u = -(\f^n)_z$ and $L_\psi v = -(\psi^n)_z$ evaluated at $t_1$ and $t_2$, respectively.  Then, using Propositions \ref{prop:powerfunctions}, \ref{prop:uhkbound}, and $\ref{prop:hkoperatordifference}$
\begin{eqnarray*}
\norm{\f^m u - \psi^m v}_{H^{k}} &\leq&  \norm{\f^m u -\f^m v}_{H^k} +\norm{\f^m v - \psi^m v}_{H^k}\\
&\leq&\paren{C\norm{\f^m-1}_{H^k}+1}\norm{u-v}_{H^k} +C' \norm{v}_{H^k} \norm{\f^m - \psi^m}_{H^k}\\
&\leq& C''\norm{\phi-\psi}_{H^k}
\end{eqnarray*}
\qed

We are now able to prove \eref{eq:mapconditionc}.  First we will establish differentiability, and then continuity, of $\Lambda[\f]$.  Let $h\neq0$.  Using Proposition $\ref{prop:hkratedifference}$,
\begin{eqnarray*}
\fl\norm{\frac{\Lambda[\phi](\cdot,t+h)-\Lambda[\phi](\cdot,t)}{h}-\f^m(\cdot,t) u(\cdot,t)}_{H^{k}} &=&\norm{\frac{\int_{t}^{t+h}\f^m(\cdot,s)u(\cdot,s) -\f^m(\cdot,t)u(\cdot,t) ds}{h}}_{H^{k}}\\
&\leq& \frac{1}{\abs{h}}\abs{\int_{t}^{t+h}\norm{\f^m(\cdot,s)u(\cdot,s) -\f^m(\cdot,t)u(\cdot,t)}_{H^k} ds}\\
&\leq &\frac{C}{\abs{h}}\abs{\int_t^{t+h} \norm{\f(\cdot,s)-\f(\cdot,t)}_{H^k}ds}
\end{eqnarray*}
Using $C^1$ continuity of $\f$ in time,  there exists $\eta$ such that 
\[
        \norm{\phi(\cdot,s)-\phi(\cdot,t)}_{H^k} \leq \eta \abs{s-t}
\]
thus, 
\[
        \norm{\frac{\Lambda[\phi](\cdot,t+h)-\Lambda[\phi](\cdot,t)}{h}-\f^m(\cdot,t) u(\cdot,t)}_{H^{k}} \leq \frac{ C \eta}{\abs{h}} \abs{\int_{t}^{t+h} \abs{s-t} ds} =C\eta \frac{\abs{h}}{2}
\]
So as $h\to 0$,
\[
        \norm{\frac{\Lambda[\phi](\cdot,t+h)-\Lambda[\phi](\cdot,t)}{h}-\f^m(\cdot,t)u(\cdot,t)}_{H^{k}}\to 0
\]
Thus it is differentiable, hence continuous, and
\begin{equation}
\label{eq:derivative}
\dt\Lambda[\f](\cdot,t) = \f(\cdot,t)^mL_{\phi(\cdot,t)}^{-1}[-(\f(\cdot,t)^n)_z]
\end{equation}

Now, let $t_{1}, t_{2}<T_1$. Using Proposition \ref{prop:hkratedifference} and \eref{eq:derivative},
\begin{eqnarray}
\norm{\dt\Lambda\phi(\cdot,t_{2})-\dt\Lambda\phi(\cdot,t_{1})}_{H^{k}}&=&\norm{\f^m(\cdot,t_2) u(\cdot,t_{2}) - \f^m(\cdot,t_1)u(\cdot,t_{1})}_{H^{1}}\nonumber\\
&\leq& C \norm{\phi(\cdot,t_2)-\f(\cdot,t_1)}_{H^k}\nonumber
\end{eqnarray}
Since $\phi$ is continuous in $t$, we are done.

We have thus established that for $T\leq T_1$, \eref{eq:mapconditionc}, and hence Proposition \ref{prop:hkmap} \eref{prop:Lambdastable} holds, where $T_1$ is determined from the values necessary for \eref{eq:mapconditiona} and \eref{eq:mapconditionb} to hold.

Now we will establish that $\Lambda$ is a contraction.  Let $t\leq T_1$.  Applying Proposition \ref{prop:hkratedifference},
\begin{eqnarray*}
\fl\norm{\Lambda\phi(\cdot,t)-\Lambda\psi(\cdot,t)}_{H^{k}}&\leq&\int_{0}^{t}\norm{\f^m L_{\phi}^{-1}[-(\phi^{n})_{z}]-\psi^m L_{\psi}^{-1}[-(\psi^{n})_{z}]}_{H^{1}}ds\\
\fl&\leq& \int_{0}^{t} C \norm{\f-\psi}_\h{k} ds \leq C t [\f-\psi]_k(t)
\end{eqnarray*}
If we now choose $t$ such that $C t<1$, say $T_2$, then we have established
\begin{equation}
        \label{eq:hkcontraction}
        [\Lambda\phi-\Lambda\psi]_k \leq \alpha[\f-\psi]_k
\end{equation}
With
\[
        \alpha=C T_2<1
\]
We have thus established that for data, $\phi_0$ satisfying conditions
\begin{equation}
        \label{hkphi0conditiona} 
        \norm{\phi_{0}-1}_{H^{k}}<R
\end{equation}
and
\begin{equation}
        \label{hkphi0conditionb}
        \norm{\frac{1}{\phi_{0}}}_{\infty}\leq\frac{1}{2\eps}
\end{equation}
for some $R>0$, $1\ge \eps>0$, and $k\ge 1$.
There will exist a $T_2>0$ such that the map $\Lambda$ will be a contraction on the space $\X_{R,\eps,T_2}^k$ proving Proposition \ref{prop:hkmap} \eref{prop:contract}.  By the contraction mapping theorem, there exists a unique fixed point in $\X_{R,\eps,T_2}^k$, proving existence and uniqueness in Theorem \ref{thm:hkfixedpoint}.  Reformulated as a local existence theorem for \eref{eq:magmaequation}, we have
\begin{thm}
\label{thm:hklocalexistence}
Let $\phi_{0}$ satisfy
\[\norm{\phi_{0}-1}_\h{k}<R\]
\[\norm{\frac{1}{\phi_{0}}}_{\infty}\leq\frac{1}{2\eps}\]
for $R>0$, $1\ge \eps>0$, and $k\ge 1$.

There exists a $T>0$ dependent on $R$, $\eps$ and $R-\norm{\phi_{0}-1}_{H^{k}}$ and a $\phi-1\in
C^{1}([0,T):H^{k}(\mathbb{R}))$ such that 
\begin{eqnarray*}
                        \phi(\cdot,t)=\phi_{0}+\int_{0}^{t}\f(\cdot,s)^m u(\cdot,s)ds\\
                        -(\f(\cdot,t)^n u(\cdot,t)_z )_z +\f(\cdot,t)^m u(\cdot,t)= -(\f(\cdot,t)^n)_z
\end{eqnarray*}
        This is a local solution to $(\ref{eq:magmaequation})$ such that
$\phi(\cdot,t)\ge\eps$ and $\norm{\phi(\cdot,t)-1}_{H^{k}}\leq R$ for $t<T$.

        Moreover, there is a maximal time of existence, $T_{\max}>0$, such that if
$T_{\max}<\infty$, then
        \begin{equation}
                \label{eq:hkblowup}
                        \lim_{t\to T_{\max}}\norm{\phi(\cdot,t)-1}_{H^{k}}+\norm{\frac{1}{\phi(\cdot,t)}}_{\infty}=\infty
        \end{equation}
\end{thm}
Proof: The first part of this theorem has already been proved.  Let us address the notion of the maximal time of existence.  Assume $T_{\max}<\infty$ and 
\[
\sup_{0\leq t <T_{\max}} \norm{\phi(\cdot,t)-1}_{H^{k}}+\norm{\frac{1}{\phi(\cdot,t)}}_{\infty}\leq K<\infty
\]
Then we have $\norm{\phi(\cdot,t)-1}_\h{k}\leq K$ and $\phi(\cdot,t)\ge  K^{-1}$.  Hence we may apply our local existence theory with $R=K+1$ and $2\eps=K^{-1}$ to get a time of size $\delta=\delta(K,k)$ over which we know we may propagate forward data satisfying these constraints.  If we then take as a new initial condition $\f_1(z)=\f(z,T_{\max}-\delta/2)$, we know it will continue up till at least time $T_{\max} +\delta/2$.  This contradicts the maximality of $T_{\max}$, proving \eref{eq:hkblowup}.\qed

The next result establishes that the solution to the initial value problem of \eref{eq:magmaequation} depends continuously on the data $\f_0$.
\begin{thm}
        \label{thm:datadependence}
        Let $\f_0$ and $\psi_0$ both satisfy
        \[\norm{\phi_{0}-1}_\h{k}<R\]
	 \[\norm{\frac{1}{\phi_{0}}}_{\infty}\leq\frac{1}{2\eps}\]
	for some $1\ge\eps>0$, $R>0$ and $k\ge 1$, and let $\f$ and $\psi$ be their respective solutions in $\X_{R,\eps,T}^k$ for some $T>0$ that both solutions are known to satisfy.  Then there exists a constant $C=C(R,\eps,k)$ such that
	\[
                \norm{\f(\cdot,t)-\psi(\cdot,t)}_\h{k}\leq  \norm{\f_0-\psi_0}_\h{k}e^{C t}
        \]
        for $t<T$.
\end{thm}
Proof: Let $t<T$.
\begin{eqnarray}
                \fl\norm{\f(\cdot,t)-\psi(\cdot,t)}_\h{k}&\leq&\norm{\f_0(\cdot) +\int_0^t \f_t(\cdot,s) ds -\psi_0(\cdot) - \int_0^t \psi_t(\cdot,s) ds}_\h{k}\nonumber\\
                \fl&\leq& \norm{\f_0-\psi_0}_\h{k} +\int_0^t \norm{\f_t(\cdot,s)-\psi_t(\cdot,s)}_\h{k}ds\nonumber
        \end{eqnarray}
Applying \eref{eq:derivative} and Proposition \ref{prop:hkratedifference}(with$ t=t_1=t_2$),
\[
        \norm{\f(\cdot,t)-\psi(\cdot,t)}_\h{k}\leq \norm{\f_0-\psi_0}_\h{k} + C\int_0^t \norm{\f(\cdot,s)-\psi(\cdot,s)}_\h{k}
\]
Gronwall's inequality then gives
\[
        \norm{\f(\cdot,t)-\psi(\cdot,t)}_\h{k}\leq  \norm{\f_0-\psi_0}_\h{k} e^{C t} \leq \norm{\f_0-\psi_0}_\h{k} e^{C T}
\]\qed
We have thus shown that the problem is locally well-posed in time for appropriate data, completing the proof of Theorem \ref{thm:hkfixedpoint}.

\section{Conservation Laws}
\label{sec:conservation}

In this section we discuss conservation laws of \eref{eq:magmaequation}. We derive a particularly useful ``energy'', which we denote by ${\cal E}$ and then prove the continuity of the mapping $\f\mapsto{\cal E}[\f]$ on a subset of $H^1$.  Finally, we show that ${\cal E}$ is locally convex about the uniform state, $\f\equiv 1$. These observations play a central role in the coming discussion of global well-posedness and stability in sections \ref{sec:smalldata} and \ref{sec:largedata}. 

A systematic search for conservation laws of \eref{eq:magmaequation} was carried out in \cite{Harris:1996xl}.  For general pairs, $(n,m)$, two conserved quantities were found. The first, corresponding to conservation of mass, is 
\begin{equation}
\label{eq:con1}
T_{1}  =  \int \phi -1 dz
\end{equation}
is conserved.  The second quantity, $T_2$ depends on the particular $(n,m)$ pair, forming the following three families
\begin{equation}
\fl T_2 = \cases{\int \frac{1}{2}\phi^{-2m}\phi_{z}^{2}+\f\log{\f} -\f +1dz & for $ n+m=1$\\
\int \frac{1}{2}\phi^{-2m}\phi_{z}^{2}-\log{\f}dz & for $n+m=2$\\
\int \frac{1}{2}\phi^{-2m}\phi_{z}^{2}+\frac{\phi^{2-n-m}-1}{(n+m-1)(n+m-2)}dz & for $n+m\neq1,2$}
\label{eq:con2}
\end{equation}
Let $\tau_2$ be the integrand of $T_2$.  Although $\tau_2$ is the same within each of these three families, the corresponding flux, $z_2$, satisfying $\partial_t \tau_2 + \partial_z z_2 = 0$ differs depending on whether or not $m=1$.  See \cite{Harris:1996xl} for the full taxonomy.

{\it  It is interesting to note that neither $T_1$ nor $T_2$ are well-defined on the spaces in the space, where local well-posedness is proved!} In particular, $\phi-1\in H^k$ does not imply $T_1[\phi]<\infty$.
Nevertheless, we have found that an appropriate linear combination of $T_1$ and $T_2$ is well-defined  and can be used in the study of global well-posedness.

This observation is linked with question of stability of the uniform state, $\phi\equiv 1$. To facilitate our study of the stability of the uniform state, we seek a linear combination of $T_1$ and $T_2$, such that: 
\begin{itemize}
\item[(a)] $\mathcal{E}[\phi] = \alpha_1 T_1[\f] + \alpha_2 T_2[\f]$, is a continuous functional of $\phi$ and well-defined on local solutions
\item[(b)] $\f\equiv1$ is a critical point of $\mathcal{E}[\phi] $ 
\item[(c)] $\mathcal{E}[\phi] $ is locally convex at $\f\equiv1$.
\end{itemize}
We find that these criteria, (a)-(c), can be satisfied if we choose $\alpha_1$ and $\alpha_2$ as follows: 
\begin{equation}
\label{eq:alpha1}
\alpha_1 = \cases{0 & for $n+m=1$\\
1 & for $n+m=2$\\
\frac{1}{n+m-1} & for $n+m\neq 1,2$
}
\end{equation}
\begin{equation}
\label{eq:alpha2}
\alpha_2 = 1
\end{equation}
Our functional then has the form\footnote{In the case $n=3$, $m=0$, $\E$ plays a role in the study of the instability of solitary waves in higher dimensions with respect to transverse perturbations  \cite{Barcilon:1989qf}.}
\begin{equation} 
\E[\f]=\int \frac{1}{2} \phi^{-2m}\phi_{z}^{2} + V_{n,m}(\f) = \alpha_1 T_1+\alpha_2 T_2
\label{eq:functional}
\end{equation}
where the potential $V_{n,m}$ function depends on the particular pair of exponents, and is given as
\begin{equation}
\fl V_{n,m}(x) = \cases{x\log{x} -x +1 & for  $n+m=1$\\
-\log{x}+x-1& for $n+m=2$\\
\frac{ x^{2-n-m}-1 + (n+m-2)(x-1)}{(n+m-1)(n+m-2)} & $n+m \neq 1,2$\\}        
\label{eq:potentials}
\end{equation}

Let us make a few remarks on $V_{n,m}$.
\begin{enumerate}
\item $V_{n,m}$ is nonnegative and $C^{\infty}([\eps,M])$, with $\eps>0$ and $M<\infty$.
\item $V_{n,m}(1)=V_{n,m}'(1)=0$, hence for $p\le2$
\[
                \frac{V_{n,m}(x)}{(x-1)^p}
\]
is also in $C^{\infty}([\eps,M])$.
\item $V_{n,m}$ is continuous as a function of $n$ and $m$.  Indeed, if $\eps = n+m-2$, then
\[
\fl\lim_{\eps\to 0} V_{n,m}(x) = \lim_{\eps\to 0} \frac{x^{-\eps}-1 + \eps(x-1)}{(\eps+1)\eps}=  \lim_{\eps\to 0}\frac{x^{-\eps}-1}{\eps} +x-1= -\log x +x -1
\]
Similarly, if $\eps = n+m-1$,
\begin{eqnarray*}
\fl \lim_{\eps\to 0} V_{n,m}(x) &=& \lim_{\eps\to 0} \frac{x^{1-\eps}-1 + (\eps-1)(x-1)}{\eps(\eps-1)}\\
&=&\lim_{\eps\to 0}\frac{x^{1-\eps}-x}{\eps}-(x-1)=x\log{x} -x +1 
\end{eqnarray*}
Hence there is really a single expression not only for the potential but also for  the functional, and the boundary cases $n+m=1$ and $n+m=2$, can be derived from it by taking limits.
\end{enumerate}

\begin{prop}
        \label{prop:functionalcontinuity}
        The functional $\mathcal{E}$ is Lipschitz continuous on the set
        \[
                \{\phi-1\in H^{1}(\mathbb{R}) :\norm{\phi-1}_\h{1}\leq R, \f\ge \eps\}
        \]
        under the $H^{1}$ norm, with Lipschitz $C=C(R,\eps,n,m)$, i.e.
        \[
        \abs{\E[\f]-\E[\psi]}\leq C \norm{\f-\psi}_\h{1}
        \]
\end{prop}
Proof: Let $\f$ and $\psi$ be two elements of this space.  Let $M=1+\sqrt{2}R$, which bounds $\norm{\f}_{L^\infty}$ by a Sobolev inequality.  Since it will be of use, let $\ell$ denote the Lipschitz constant for $V_{n,m}(x)/(x-1)^2$ on the set $[\eps,M]$ and let $N$ denote this quotient's $L^\infty$ norm on this set.  Then, 
\begin{eqnarray*}
\fl\abs{\mathcal{E}[\f]-\mathcal{E}[\psi]}&\leq & \frac{1}{2} \int \abs{\frac{\f_z^2}{\f^{2m}}-\frac{\psi_z^2}{\psi^{2m}}}+\int \abs{V_{n,m}(\f)-V_{n,m}(\psi)}\\
              \fl&\leq &\frac{1}{2}\int \abs{\frac{\f_z^2 -\psi_z^2}{\f^{2m}}} + \abs{\psi_z^2\paren{\frac{1}{\f^{2m}}-\frac{1}{\psi^{2m}}}}\\
               \fl& &+\int \abs{\frac{V_{n,m}(\f)}{(\f-1)^2}(\f-1)^2 - \frac{V_{n,m}(\psi)}{(\psi-1)^2}(\psi-1)^2 } \\
                \fl&\leq & \frac{1}{2}\paren{\norm{\frac{1}{\f^{2m}}}_\infty \int \abs{\f_z-\psi_z}\abs{\f_z+\psi_z} + \norm{\frac{\f^{2m}-\psi^{2m}}{\f^{2m}\psi^{2m}}}_\infty \int \psi_z^2}\\
                \fl& &+\int \abs{\frac{V_{n,m}(\f)}{(\f-1)^2}}\abs{(\f-1)^2-(\psi-1)^2} +\int \abs{(\psi-1)^2}\abs{V_{n,m}(\f)-V_{n,m}(\psi)}\\
                \fl&\leq& C\norm{\f_z-\psi_z}_{L^2} + C' \norm{\f_z-\psi_z}_\h{1} +C''\norm{\f-\psi}_{L^2}+C'''\norm{\f-\psi}_{L^2}\\
                \fl&\leq&  \tilde{C} \norm{\f-\psi}_\h{1}
\end{eqnarray*}\qed

The following result shows that $\f_*\equiv1$ is a critical point of ${\cal E}$ and  for any $(n,m)$,  ${\cal E}$ is locally convex near  $\f_*$. 
\begin{thm}
        \label{thm:variationalbound}
        Let $p\in \h{1}(\R)$.  Then
        \[
                \mathcal{E}[1+p]=\frac{1}{2}\norm{p}_\h{1}^2 + \Or(\norm{p}_\h{1}^3)
        \]
        as $\norm{p}_\h{1}\to 0$.
\end{thm}
Proof:  Each of the cases of \eref{eq:potentials} must be examined separately, however the scheme is universal.  We present the proof for $n+m>2$.  Given $p$, let $f=1+p$.  
The first variation of $\mathcal{E}[\phi]$ is
\begin{eqnarray*}
\fl\langle\delta \mathcal{E}[1+p], p\rangle&=&\int \paren{\frac{1}{n+m-1}-\frac{1}{n+m-1} \frac{1}{f^{n+m-1}}-m\frac{f_z^2}{f^{1+2m}}}p+\frac{ f_z }{ f^{2m} }p'\\
\fl               &=&\int \paren{\frac{1}{n+m-1}- \frac{1}{n+m-1}\frac{1}{f^{n+m-1}}+m\frac{f_z^2}{f^{1+2m}} -\frac{f_\zz}{f^{2m}}  }p
\end{eqnarray*}
And the second variation is
\[
\fl\langle\delta^{2}\mathcal{E}[1+p]p,p\rangle=\int \paren{\frac{1}{f^{n+m}}  -m(1+2m)\frac{f_z^2}{f^{2+2m}} + 2m\frac{f_\zz}{f^{1+2m}}  }{p}^{2} +\paren{\frac{1}{f^{2m}}}{p'}^{2}
\]
Thus, taking a Taylor Expansion about $p=0$,
\[
\fl\mathcal{E}[1+p] = \mathcal{E}[1] +\langle\delta\mathcal{E}[1],p\rangle +\frac{1}{2} \langle\delta^2\mathcal{E}[1]p,p\rangle+\Or(\norm{p}_\h{1}^3)= \frac{1}{2}\norm{p}_\h{1}^2 + \Or(\norm{p}_\h{1}^3)
\]
\qed

In \sref{sec:smalldata} we shall prove conservation of ${\cal E}$; see Propositions \ref{prop:energyconservation} and \ref{prop:smallh1conservation}.
 
\section{Data near the uniform state: Global existence and Lyapunov stability}
\label{sec:smalldata}
We now prove \underline{global} well-posedness for data sufficiently close to the uniform state, $\f\equiv1$.  Furthermore, we show Lyapunov stability of the uniform state.  An outline of our strategy is
\begin{enumerate}
\item Establish conservation of $\mathcal{E}[\f]$ for solutions of $H^2$ spatial regularity.
\item Construct $H^1$ \emph{a priori} estimates  using conservation of $\E[\f]$ and Theorem \ref{thm:variationalbound} when $\f$ is of $H^2$ spatial regularity
\item Construct $H^2$ \emph{a priori} estimates by deriving a growth estimate on $\norm{\f_{zz}}_{L^{2}}$ for $\h{2}$ solutions to \eref{eq:magmaequation} 
\item Use these bounds to get $\h{2}$ global existence
\item Exhibit stability in the $\h{1}$ norm for $\h{2}$ solutions with $\norm{\f_0-1}_\h{1}$ sufficiently small
\item Establish that $\E$ is also conserved for $\h{1}$ solutions by approximating $\h{1}$ data by $\h{2}$ data and taking appropriate limits
\item Prove $\h{1}$ global existence by using the $\h{2}$ argument, with the $\h{2}$ norm replaced with $\h{1}$ now that we have $\mathcal{E}[\f]=\mathcal{E}[\f_0]$
\end{enumerate}

\begin{prop}
\label{prop:energyconservation}
Assume $\f-1 \in C^1([0,T):\h{2}(\R))$ is a solution to \eref{eq:magmaequation}, then $\mathcal{E}[\f]$ is conserved.
\end{prop}

Proof:  Since $\f$ is of $H^2$ spatial regularity, we may expand \eref{eq:magmaequation} to 
\[
        \f_t +(\f^n + m\f^{n-m-1}\f_t\f_z -\f^{n-m}\f_{tz})_z=0
\]
Multiplying by 
\[
        \frac{1}{n+m-1}\paren{1-\f^{1-m-n} }
\]
for $n+m\neq 1$ and $\log \f$ for $n+m=1$, one will find, after integrating by parts, the expression may be rearranged into
\[
\dt e +\dz f = 0
\]
where, for $n+m\neq 1,2$, 
\begin{eqnarray*}
\fl e &=& \frac{1}{2}\frac{\f_z^2}{\f^{2m}} + \frac{\f^{2-m-n}-1 + (n+m-2)\paren{\f-1}}{(n+m-1)(n+m-2)}\\
\fl f &=& \frac{1}{n+m-1}\paren{\f^n + m\f^{n-m-1}\f_t\f_z -\f^{n-m}\f_{tz}-\frac{n}{1-m}\f^{1-m} -m\f^{2m}\f_z\f_t+\f^{1-2m}\f_{zt}}
\end{eqnarray*}
and $\int e = \E$.  Analogous expressions hold in the other cases.  Note that the factor by which we multiply is continuous in the limit of $n+m=1$.\qed

\begin{prop}
        \label{prop:secondderivativebound} Assume $\f-1 \in C^1([0,T):\h{2}(\R))$ is a solution to \eref{eq:magmaequation}, and there exist constants $R, \eps$, such that
\[
\norm{\f(\cdot,t)-1}\leq R
\]
\[
\norm{\frac{1}{\f(\cdot,t)}}_\infty \leq \frac{1}{\eps}
\]
for $t< T$.
Then there exist constants $c_0=c_0(R,\eps,n,m)$ and $c_1=c_1(R,\eps,n,m)$ such that
        \[
                \frac{d}{dt}\norm{\phi_{zz}(\cdot,t)}_{2}\leq c_0 +c_1\norm{\phi_{zz}(\cdot,t)}_{2}
        \]
\end{prop}
Proof:  Expanding out  \Eref{eq:magmaequation},
\[
\fl\f_t + n \f^{n-1}\f_z +m(n-m-1)\f^{n-m-1}\f_z^2\f_t -(n-2m)\f^{n-m-1}\f_z \f_{zt} + m\f^{n-m-1}\f_t\f_\zz -\f^{n-m}\f_t\f_{zzt}=0
\]
Which we rewrite as
\[
\fl\f_{zzt}= \frac{\f_t}{\f^{n-m}} +n \frac{\f_z}{\f^{1-m}}+m(n-m-1)\frac{\f_z^2\f_t}{\f^2}-(n-2m)\frac{\f_z\f_{zt}}{\f} + m \frac{\f_t\f_\zz}{\f}
\]
Letting $u$ satisfy $L_\f u = -(\f^n)_z$, $\f_t = \f^m u$, and
\[
\fl \f_{zzt}= \frac{u}{\f^{n-2m}} +n \frac{\f_z}{\f^{1-m}}+m(n-m-1)\frac{\f_z^2u}{\f^{2-m}}-m(n-2m)\frac{\f_z^2 u}{\f^{2-m}} -(n-2m)\frac{\f_z u_z}{\f^{1-m}}+ m \frac{u\f_\zz}{\f^{1-m}}
\]
Now, let us study the evolution of the $\norm{\f_\zz}$ in time.
\begin{eqnarray*}
\fl\frac{d}{dt}\frac{1}{2}\norm{\f_\zz}_2^{2}&=&\int \phi_{zz}\phi_{zzt}=\int \f_\zz \bigg\{\frac{u}{\f^{n-2m}} +n \frac{\f_z}{\f^{1-m}}+m(m-1)\frac{\f_z^2u}{\f^{2-m}} -(n-2m)\frac{\f_z u_z}{\f^{1-m}}+ m \frac{u\f_\zz}{\f^{1-m}}\bigg\}\\
                \fl&\leq& \lefteqn{\norm{\frac{1}{\f^{n-2m}}}_\infty\norm{\f_\zz}_2\norm{u}_2 + \abs{n}\norm{\frac{1}{\f^{1-m}}}_\infty \norm{\f_\zz}_2\norm{\f_z}_2} \\
                \fl& &+ \abs{m(1-m)}\norm{\frac{1}{\f^{2-m}}}_\infty\norm{\f_\zz}_2\norm{\f_z^2u}_2+\abs{n-2m}\norm{\frac{1}{\f^{1-m}}}_\infty\norm{\f_\zz}_2\norm{\f_z u_z}_2\\
                \fl& &+ \abs{m}\norm{\frac{u}{\f^{1-m}}}_\infty \norm{\f_\zz}_2^2\\
                \fl&\leq& C \norm{u}_2\norm{\f_\zz}_2 + C'\norm{\f_\zz}_2+C''\norm{\f_z}_\infty\norm{u}_\infty\norm{\f_\zz}_2\\
                & &+C'''\norm{\f_z}_\infty\norm{u_z}_2\norm{\f_\zz}_2+C''''\norm{u}_\infty\norm{\f_\zz}_2^2
               \end{eqnarray*}
Recall Proposition $\ref{prop:uhkbound}$ that there exists a constant depending on $R$, $\eps$, and $k=1$ such that
\[
        \norm{u}_\h{1} \leq \tilde{C}
\]
Also, an application of Sobolev's inequality gives the bound
\[
        \norm{\f_z}_\infty \leq \sqrt{2}\norm{\f_z}_2\norm{\f_\zz}_2\leq \sqrt{2} R \norm{\f_\zz}_2
\]
Using these bounds,
\[                
\frac{d}{dt}\frac{1}{2}\norm{\f_\zz}_2^{2}\leq c_0 \norm{\f_\zz}_{L^2} + c_1\norm{\f_zz}_{L^2}^2
\]
Therefore,
\[
        \frac{d}{dt}\norm{\f_\zz}_2\leq c_0 + c_1 \norm{\f_\zz}_2
\]
\qed

\begin{cor}
        \label{cor:secondderivativebound}
        Under the assumptions of Lemma $\ref{prop:secondderivativebound}$,
        \[
                \norm{\dzz\f(\cdot,t)}_2 \leq c_0 t e^{c_1 t} + \norm{\dzz\f_0}_2 e^{c_1 t}
        \]
\end{cor}
Proof: Apply Gronwall's inequality.\qed

We note that no assumption on the size of the data was used in proving Propositions \ref{prop:energyconservation} and \ref{prop:secondderivativebound}; they will also be used in the case of "large" data in \sref{sec:largedata}.

\begin{thm}(Global existence in $\h{2}$ and stabilty near the uniform state in $\h{1}$)
        \label{thm:smallh2globalexistence}
        
        There exists $\eta_c=\eta_c(n,m)>0$ such that for any $\eta<\eta_c$ there exists $\delta=\delta(\eta,n,m)$ such that for any $\f_0$ satisfying
        \begin{eqnarray}
                \label{eq:smallh2globalconditiona}
                \norm{\phi_{0} -1}_\h{2}<\infty\\
                \label{eq:smallh2globalconditionb}
                \norm{\phi_0-1}_\h{1}\leq\delta
        \end{eqnarray}        
       there exists a unique $\f$, $\f-1 \in C^{1}([0,\infty):H^{2}(\R))$ such that solves \eref{eq:magmaequation} with $\f(\cdot,0)=\f_0$.
Moreover,        
        \[
        	\norm{\f(\cdot,t)-1}_\h{1}\leq \eta
        \]
        for all time.
        \end{thm} 
Proof:  First, let us construct $\eta_c$.  Proposition \ref{prop:energyconservation} ensures that $\mathcal{E}$ is conserved for such a solution.  From Theorem \ref{thm:variationalbound}, there exist constants $C$ and $D$ such that for $\norm{\f-1}_\h{1}\le\sqrt{2}/4$,
\begin{equation}
\fl \frac{1}{2}\norm{\f-1}_\h{1}^2 -C \norm{\f-1}_\h{1}^3\le \E[\f]\le\frac{1}{2}\norm{\f-1}_\h{1}^2 +D \norm{\f-1}_\h{1}^3
\label{eq:variationalinequality}
\end{equation}
The peak of $1/2 x^2 - C x^3$ occurs at $x=1/(3C)$, and we set $\eta_c = \min\{1/(3C),\sqrt{2}/4\}$.  Let $E_c =1/2 \eta_c^2 - C \eta_c^3$.  Given $\eta<\eta_c$, let $E = 1/2 \eta^2 - C \eta^3$ and let $\delta$ solve $E=1/2 \delta^2 + D \delta^3$.  As indicated in \fref{fig:variationalbound}, this construction forces
\[
	\delta<\eta<\eta_c\leq \sqrt{2}/4
\]
and for $\norm{\f-1}_\h{1}\leq \delta$
\[
\E[\f]\leq E<E_c
\]

\begin{figure}
        \begin{center}
                \caption{The two curves provide bounds on $\E{\f}$ for $\norm{\f-1}_\h{1}\leq \sqrt{2}/4$.  In this picture we have assumed that the peak of the lower bound is less than $\sqrt{2}/4$, hence $\eta_c$ is the peak.  Taking $\eta<\eta_c$, we are then able to constrain the energy by choosing $\norm{\f_0-1}_\h{1}\leq \delta$ to $\E[\f(\cdot,t)]\leq E$.  This assures us that $\norm{\f(\cdot,t)-1}_\h{1}\leq\eta$.}
                \label{fig:variationalbound}
        \end{center}
\end{figure}

Now assume that $\norm{\f_0-1}_\h{1}\leq \delta$ in addition to having $\norm{\f_0-1}_\h{2}<\infty$. From Theorem \ref{thm:hklocalexistence}, we know there exists $T>0$ such that $\f-1 \in C^1([0,T):\h{2}(\R))$ that solves \eref{eq:magmaequation} with $\f_0$ as the initial condition.

Suppose for some time $\overline{T}$,  $\phi-1$ ceases to be in $C^{1}([0,\overline{T}):\h{2}(\R))$.  Furthermore, let $\overline{T}$ be the minimal such value.  If $\overline{T}=\infty$, then we are done, so we may take $\overline{T}<\infty$. 

For $t<\overline{T}$,  $\phi-1\in C^{1}([0,t]:\h{2}(\R))$; hence it is a strong solution to
\eref{eq:magmaequation} and $\E[\f]$ is conserved.  This implies
\[
	\norm{\f(\cdot,t)-1}_\h{1}\le \eta
\]
for $t<\overline{T}$.

Suppose not.  Then there exists $t<\overline{T}$ for which the reverse holds, and by the continuity of $\f$, there exists a minimal time $t_1$ such that $\eta< \norm{\f(\cdot,t_1)}_\h{1} <\eta_c$.  However,
\[
	\norm{\f(\cdot,t_1)}_\h{1} <\eps_c\leq \frac{\sqrt{2}}{4}
\]
So \eref{eq:variationalinequality} holds with the same constants $C$ and $D$, but this implies,
\[
\fl\E[\f(\cdot,t_1)]\ge\frac{1}{2}\norm{\f(\cdot,t_1)-1}_\h{1}^2 -C \norm{\f(\cdot,t_1)-1}_\h{1}^3 > \frac{1}{2}\eta^2 -C \eta^3 = E \ge \E[\f_0]
\]
contradicting conservation of $\E$.

By construction $\eta<\eta_c\leq \sqrt{2}/4$, so   $\f\geq \frac{1}{2}$.  Applying Proposition \ref{prop:secondderivativebound}, with $R=\eta_c$ and $\eps=\frac{1}{2}$, we have that for $t<\overline{T}$, 
\begin{eqnarray}
\norm{\f(\cdot,t)-1}_\h{2}&\leq &\eta + c_0 t e^{c_1 t} + \norm{\f(\cdot,0)_zz}e^{c_1 t} \nonumber \\
&\leq&   \frac{\sqrt{2}}{4} + c_0 \overline{T} e^{c_1 \overline{T}} + \norm{\f(\cdot,0)_\zz}e^{c_1 \overline{T}}=R\\
\norm{\frac{1}{\f}}_\infty \leq 2
\end{eqnarray}
Hence for all $t<\overline{T}$,
\[
\norm{\phi(\cdot,t)}_{H^{2}}+\norm{\frac{1}{\phi(\cdot,t)}}_{\infty}\leq R+2<\infty
\]
and this sum does not go to infinity, as required by \eref{eq:hkblowup}, for the solution to have a finite time of existence.  Therefore, the solution persists for all time.   Additionally, we have
\[
\norm{\f(\cdot,t)-1}_\h{1}\leq \eta
\]
\qed

\begin{prop}
        \label{prop:smallh1conservation}
        Let $\phi_{0}$ satisfy \eref{eq:smallh2globalconditionb} and let $\phi(z,t;\phi_{0})-1\in C^{1}([0,T):H^{1}(\R))$ be the solution of the initial value problem of \eref{eq:magmaequation} with this data.  Then $\mathcal{E}[\phi(\cdot,t;\phi_{0})]$ is conserved on the interval $[0,T)$.
\end{prop}
Note that $T>0$ by the assumptions on the data and Theorem \ref{thm:hklocalexistence}.

Proof: Let $\{\phi_{0,n}\}$ be a sequence satisfying
        \begin{eqnarray*}
                \norm{\phi_{0}-1}_\h{2}&<\infty\\
                \norm{\phi_{0,n}-1}_\h{1}&\leq\delta
        \end{eqnarray*}
in addition to assuming that
\begin{equation}
        \label{eq:limitcondition}
        \lim_{n\to\infty}\norm{\phi_{0}-\phi_{0,n}}_\h{1}=0
\end{equation}
Here, we have used the density of $H^{2}$ within $H^{1}$ to find such functions. 

Since $\phi_{0,n}$ have global solutions by Theorem \ref{thm:smallh2globalexistence},   we can use Theorem \ref{thm:datadependence} on the space $\X_{\eps,\frac{1}{4},\infty}^1$, to get
\[
        \norm{\phi(\cdot,t;\phi_{0,m})-\phi(\cdot,t;\phi_{0,n})}_{H^{1}}\leq\norm{\phi_{0,m}-\phi_{0,n}}_{H^{1}}e^{C t}
\]
Letting $m\to\infty$, for $t<\infty$,
\begin{equation}
        \label{eq:limitdifference}
        \norm{\phi(\cdot,t;\phi_{0})-\phi(\cdot,t;\phi_{0,n})}_{H^{1}}\leq\norm{\phi_{0}-\phi_{0,n}}_{H^{1}}e^{C T}
\end{equation}
Now, consider the change in the functional $\mathcal{E}$, over time,
\begin{eqnarray*}
 \abs{\mathcal{E}[\phi(\cdot,t;\phi_{0})]-\mathcal{E}[\phi_{0}]}\leq \lefteqn{ \abs{\mathcal{E}[\phi(\cdot,t;\phi_{0})]-\mathcal{E}[\phi(\cdot,t;\phi_{0,n})]} }\\
& & +\abs{\mathcal{E}[\phi(\cdot,t;\phi_{0,n})]-\mathcal{E}[\phi_{0,n}]}+\abs{\mathcal{E}[\phi_{0,n}]-\mathcal{E}[\phi_{0}]}
\end{eqnarray*}
Since $\phi(z,t;\phi_{0,n})$ are global strong solutions, $\E[\phi(z,t;\phi_{0,n})]$
is conserved.  Applying Proposition \ref{prop:functionalcontinuity} with the set
\[
\{\f-1 \in \h{1}(\R) : \norm{\f-1}_\h{1}\leq \eps, \f\geq \frac{1}{2}\}
\]
we get
\[
        \abs{\E[\phi(\cdot,t;\phi_{0})]-\E[\phi_{0}]}\leq C(\norm{\phi(\cdot,t;\phi_{0})-\phi(\cdot,t;\phi_{0,n})}_{H^{1}}+\norm{\phi_{0}-\phi_{0,n}}_{H^{1}})
\]
Now, using \eref{eq:limitdifference},
\[
        \abs{\E[\phi(\cdot,t;\phi_{0})]-\E[\phi_{0}]}\leq C'\norm{\phi_{0}-\phi_{0,n}}_{H^{1}}
\]
Letting $n\to\infty$, we have for $t<T$,
\[
        \E[\phi(\cdot,t;\phi_{0})]=\E[\phi_{0}]
\]\qed

\begin{cor}
        \label{cor:smallh1globalexistence}
Theorem \ref{thm:smallh2globalexistence} holds for $\h{1}$. 
\end{cor} 

While we have not proved global existence in $C^1([0,T):\h{k}(\R))$ for $k>2$, the results of this section still yield a stability result in these smoother spaces.  We know we have a local solution in these spaces from Theorem \ref{thm:hklocalexistence}.  Proposition \ref{prop:energyconservation} and \ref{prop:smallh1conservation} establish conservation of $\mathcal{E}$ for solutions in $C^1([0,T):\h{k}(\R))$, with $k\geq 1$.  Following the same argument as used in theorem \ref{thm:smallh2globalexistence}, we thus have
\begin{cor}(Nonlinear Stability of the Uniform State)
\label{cor:nonlinearstability}
There exists $\eps_c>0$ such that for any $\eps< \eps_c$ there exists $\delta$ such that if $\norm{\f_0-1}_\h{1}\le\delta$, then the solution to \eref{eq:magmaequation}, $\f\in C^1([0,T):\h{k}(\R))$, $k\ge 1$, $T>0$, allowing $T=\infty$, with this initial condition satisfies
\[
\norm{\f(\cdot,t)-1}_\h{1}\le\eps
\]
for $t<T$.
\end{cor}

\section{Global Well-Posedness for (almost) arbitrary data}
\label{sec:largedata}
We now turn our attention to the case of data that may not be close to the uniform state in the $\h{1}$ norm.  While the data must still be bounded away from zero, we are able to show global well-posedness for data of arbitrarily large size in the $\h{1}$ norm, provided we satisfy the following three conditions on $(n,m)$.
\numparts
\begin{eqnarray}
\label{eq:nmconditiona}
                n+m\geq 3\\
\label{eq:nmconditionb}
                n+3m\geq 4\\
\label{eq:nmconditionc}
                n+3m < 6
\end{eqnarray}
\endnumparts  
This is the shaded region in \fref{fig:parameterplot}.  The requirement that $m>0$ for smaller values of $n$ is consistent with our suspicion that having a variable bulk viscosity regularizes the problem, as this requires progressively more work to be done to expel fluid from the pores as the porosity becomes smaller.  For the remainder of this section, we assume these conditions to hold.

We must derive certain conserved quantities in this special case.  Recalling \eref{eq:functional}, we write
\begin{equation}
        \label{eq:potential1}
        g_r(x)=r x^{r+1}-(r+1) x^r +1
\end{equation}
With $r=n+m-2$, we may write
\[
        \E[\f] = \int \frac{1}{2} \frac{\f_z^2}{\f^{2m}} + \frac{1}{r(r+1)}\frac{g_{r}(\f)}{\f^{r}}
\]
\begin{lem}
        \label{lem:largedatapotentialbound}
        For $r\geq 1$, $x\geq 0$,
        \[
                g_{r}(x) \geq (x-1)^2
        \]
\end{lem}

\begin{prop}
        \label{prop:conservedupperbound}
        Assuming $\E[\f]<\infty $ and  $\phi> 0$, there exists $M=M(\E[\f],n,m)<\infty$, such that 
        \[
        \norm{\phi}_\infty\leq M
        \]
\end{prop}
Proof:
\begin{eqnarray*}
                \frac{1}{2}(\phi(z)-1)^{2}&=&\int_{-\infty}^{z}(\phi(z)-1)\phi_{z}(z)dz\leq \int_{-\infty}^{\infty}\abs{\phi(z)-1}\abs{\phi_{z}}dz\\
               & =&\int \frac{\abs{\f_z}}{\f^m}\frac{\abs{\f-1}}{\f^{(n+m-2)/2}}\f^{m+(n+m-2)/2}\\
               &\leq& \norm{\f}_\infty^{(n+3m-2)/2} \norm{\frac{\f_z}{\f^m}}_2 \norm{\frac{\f-1}{\f^{(n+m-2)/2}}}_2\\
\end{eqnarray*}
We are able to move the exponent outside the $L^\infty$ norm since $n+3m-2>0$ by \eref{eq:nmconditionb}.  Then employing \eref{eq:nmconditiona} to use Lemma \ref{lem:largedatapotentialbound}, 
\begin{eqnarray*}
               \frac{1}{2}(\phi(z)-1)^{2}&\leq &  \norm{\f}_\infty^{(n+3m-2)/2} \sqrt{2\E[\f]} \norm{\sqrt{\frac{g_r(\f)}{\f^{r}}}}_2\\
               &\leq & \norm{\f}_\infty^{(n+3m-2)/2} \sqrt{2 r (r+1)}\E[\f]
\end{eqnarray*}
Hence,
\[
\norm{\f}_\infty -1 \leq \norm{\f-1}_\infty \leq \sqrt{2}  \norm{\f}_\infty^{(n+3m-2)/4} \paren{2 r (r+1)}^{1/4}\sqrt{\E[\f]}
\]
or
\[
\norm{\f}_\infty^{1-(n+3m-2)/4}\leq \sqrt{2} \paren{2 r(r+1)}^{1/4}\sqrt{\E[\f]}+1
\]
which ensures an upper bound so long as $1-(n+3m-2)/4>0$, which is precisely \eref{eq:nmconditionc}.
\qed

This will ensure $\f$ is pointwise bounded from above in terms of the data through conservation of $\E$.  Let us define the upper bound functional that corresponds to this maximum as 
\begin{equation}
        \label{eq:upperboundfunctional}
        \mathcal{U}[\f] = \sqrt[1-(n+3m-2)/4]{\sqrt{2} \paren{2r(r+1)}^{1/4}\sqrt{\E[\f]}+1}
\end{equation}

\begin{prop}
        \label{prop:conservedlowerbound}
        Assuming $\mathcal{E}[\f]<\infty$ and $\f>0$, there exists $\eps=\eps(\E[\f],n,m)>0$ such that
        \[
                \norm{\frac{1}{\f}}_\infty \leq \frac{1}{\eps}
        \]
\end{prop}
Proof:
\begin{eqnarray*}
\log \frac{1}{\f(z)} &=&  - \int_{-\infty}^z \frac{\f_z}{\f} dz = - \int_{-\infty}^{z} \frac{\f_z}{\f^{m+r/2}} \frac{\f^{m+r/2}}{\f} dz\\
&=& - \int_{-\infty}^{z} \paren{\frac{\f_z}{\f}\frac{1-\f}{\f^{r/2}} +\frac{\f_z}{\f^{m+r/2-1}} }\f^{m+r/2-1} dz\\
&\leq& \norm{\f^{m+r/2-1}}_\infty \norm{\frac{\f_z}{\f^m}}_2\norm{\frac{1-\f}{\f^{r/2}}}_2 - \int_{-\infty}^z \f_z dz\\
&\leq&  \norm{\f^{m+r/2-1}}_\infty \sqrt{2 r (r+1)}\mathcal{E}[\f] +1\\
&\leq& \U[\f]^{(n+3m-4)/2}\sqrt{2 r (r+1)}\mathcal{E}[\f] +1= \log\frac{1}{\eps}
\end{eqnarray*}
where we have explicitly used \eref{eq:nmconditionb} to move the exponent outside of the $L^\infty$ norm, and we have implicitly used the other two conditions along with this one, in order to make use of the conserved upper bound,$\U$.
\qed
We denote the corresponding functional  for $\eps$ as $\mathcal{L}$, given as
\begin{equation}
        \label{eq:lowerboundfunctional}
        \mathcal{L}[\f]=\exp\left\{-\U[\f]^{(n+3m-4)/2}\sqrt{2 r (r+1)}\E[\f] -1\right\}
\end{equation}
This will ensure a conserved lower bound exists by conservation of $\E$.

\begin{prop}
        \label{prop:conservedh1bound}
        Assuming $\mathcal{E}[\f]<\infty$ and $\phi> 0$, there exists $R=R(\E[\f],n,m)<\infty$ such that \[
        \norm{\phi-1}_{H^{1}}\leq R
        \]
\end{prop}
Proof:  We divide our proof into two cases $(i): m\geq 0$ and $(ii): m<0$.  For $m\geq 0$,
        \begin{eqnarray*}
                \E[\f]&\geq &\int \frac{1}{2 \U[\f]^{2m}} \f_z^2 + \frac{1}{\U[\f]^{n+m-2} r(r+1)}(\f-1)^2 dz\\
        &\geq & \min\left\{ \frac{1}{2 \U[\f]^{2m}},\frac{1}{\U[\f]^{n+m-2} r(r+1)}\right\} \norm{\f-1}_{\h{1}}^2
        \end{eqnarray*}
and we have the constant $R$.

For $m<0$, 
        \begin{eqnarray*}
                \E[\f]&\geq &\int \frac{\mathcal{L}[\f]^{-2m}}{2 } \f_z^2 + \frac{1}{\U[\f]^{n+m-2} r(r+1)}(\f-1)^2 dz\\
        &\geq & \min\left\{ \frac{\mathcal{L}[\f]^{-2m}}{2 },\frac{1}{\U[\f]^{n+m-2} r(r+1)}\right\} \norm{\f-1}_{\h{1}}^2
         \end{eqnarray*}

\qed
Again, let us define a corresponding functional that provides a conserved bound on the $\h{1}$ norm.
\begin{equation}
        \label{eq:h1functional}
        \fl\mathcal{R}[\f] =\cases{ \sqrt{\E[\f]\min\left\{ \frac{1}{2 \mathcal{U}[\f]^{2m}},\frac{1}{\mathcal{U}[\f]^{n+m-2} r(r+1)}\right\}^{-1}  } & for $m\geq0$\\
         \sqrt{\E[\f]\min\left\{ \frac{\mathcal{L}[\f]^{-2m}}{2 },\frac{1}{\U[\f]^{n+m-2} r(r+1)}\right\}^{-1}}& for $m<0$ }
\end{equation}

Now that we have \emph{a priori} bounds on $\f$ pointwise from above and below, together with a bound on the $\h{1}$ norm, we may prove global existence.  Thus we have,
\begin{thm}
        \label{thm:largeh2globalexistence}
        Let $\phi_{0}$ satisfy,
        \begin{eqnarray}
                \label{eq:largeh2globalconditiona}
                \norm{\phi_{0}-1}_{H^{2}}<\infty\\
                \label{eq:largeh2globalconditionb}
                \norm{\frac{1}{\phi_{0}}}_{\infty}<\infty
        \end{eqnarray}
        Then there exists a unique $\phi$,  $\f-1 \in C^{1}([0,\infty):H^{2}(\R))$ such that solves \eref{eq:magmaequation} with $\f(\cdot,0)=\f_0$.  Moreover, $\norm{1/\f}_\infty<\infty$ for all time.
\end{thm} 
Proof: This proof is much the same as in the small data case, with the pointwise lower and upper pointwise bounds coming from the functionals $\Low[\f]$ and $\U[\f]$, together with the norm bound $\cal{R}[\f]$, and an application of Proposition \ref{prop:secondderivativebound}.

We may prove global existence in $\h{1}$ as in the small data case by first showing that $\E$ is conserved for $\h{1}$ solutions by taking a limit of $\h{2}$ solutions, 

\begin{prop}
        \label{prop:largeh1conservation}
        Let $\phi_{0}$ satisfy 
        \begin{eqnarray}
                \label{eq:h1conservedconditiona}
                \norm{\phi_{0}-1}_{H^{1}}<\infty\\
                \label{eq:h1conservedconditionb}
                \norm{\frac{1}{\phi_{0}}}_{\infty}<\infty
        \end{eqnarray}
        and let $\phi(z,t;\phi_{0})-1\in C^{1}([0,T):H^{1}(\R))$ be the solution with data $\f_0$.  Then $\mathcal{E}[\phi(\cdot,t;\phi_{0})]$ is conserved on the interval $[0,T)$.
\end{prop}
Proof: This proof is very similar to the small data case.  Let $\eps$ be the conserved lower bound, given by $\mathcal{L}[\f_0]$ and let $R=\mathcal{R}[\f_0]$ be the conserved upper bound.  Requiring our strongly convergent sequence $\{\phi_{0,n}\}$ to satisfy
        \begin{eqnarray*}
                \norm{\phi_{0,n}-1}_{H^{2}}<\infty\\
                \norm{\frac{1}{\phi_{0,n}}}_\infty<\infty\\
                \mathcal{R}[\f_{0,n}] \leq R\\
                \mathcal{L}[\f_{0,n}]  \geq \eps 
        \end{eqnarray*}
we proceed as in the small data case almost verbatim.\qed

\begin{cor}
        \label{cor:largeh1globalexistence}
Theorem \ref{thm:largeh2globalexistence} holds for $\h{1}$. 
\end{cor} 

\begin{cor}
\label{cor:degeneracy}
For data satisfying \eref{eq:h1conservedconditiona} and \eref{eq:h1conservedconditionb}, global solutions exist in $C^{1}([0,\infty): H^{1}(\R))$ even when $m=0$ provided $4\le n<6$.
\end{cor}

We have thus established global well-posedness for data that may not be close to the uniform state in the $\h{1}$ norm, provided we satisfy (\ref{eq:nmconditiona} - \ref{eq:nmconditionc}).

\section{Discussion}
\label{sec:discussion}
Our definition of solutions of \eref{eq:magmaequation} and our proofs of existence, both locally and globally in time, depend upon ensuring $\phi > 0$.  It is this condition, and the use of Sobolev type inequalities to adhere to it, that limit our global existence proofs to the $(n,m)$ pairs in the set defined by (\ref{eq:nmconditiona} - \ref{eq:nmconditionc}).  However, numerical experiments indicate that we should expect a global lower bound on $\phi$ to exist at least for the values of $[2,5]\times[0,1]$.  On the other hand, solutions have been proposed in \cite{Nakayama:1991ud, Nakayama:1994cq,Takahashi:1992lp,Rabinowicz:2002bj} which are zero on sets of measure zero.  This leaves us with three questions:
\begin{itemize}
  \item Why do solutions that are initially uniformly bounded below away from zero remain above zero?
  \item How might a global existence proof be achieved for other choices of exponents $n$ and $m$?
  \item What is to be made of these functions that go to zero on sets of measure zero?
\end{itemize}
We offer partial answers to these questions and suggest directions for future work in this final section.

\subsection{Importance of Nonlinear Dispersion for Positivity}
The appearance of a degenerate nonlinearity in the dispersive term appears to be key to the positivity of the solution.  We saw this in \Sref{sec:largedata}, where the appearance of terms $\propto \phi^{2-n-m}$ in the invariants prevented $\phi$ from going to zero.  These terms only appear in the conserved integrals because of their presence in the underlying equation.  There are physical reasons to suspect the benefit of a variable bulk viscosity in this model, consistent with \eref{eq:nmconditionb}; by increasing $m$ above zero, we gain global existence for a continuum of values of $n$, for which we could not otherwise succeed.  But while having nonzero $m$ may permit global existence for certain $n$, it is by no means required, as noted in Corollary \ref{cor:degeneracy}.

Rather than continue to look at the invariants, let us examine the evolution itself.  Comparing 
\begin{equation}
\label{eq:wellposedmagma}
	\f_t + (\f^4)_z -(\f^4 \f_{zt})_z=0
\end{equation}
the magma equation with $n=4$ and $m=0$, shown to be globally well posed for arbitrary data in \Sref{sec:largedata}, with a generalized BBM (gBBM) equation,
\begin{equation}
\label{eq:bbmcomparison}
	u_t +(u^4)_x -u_{xxt}=0
\end{equation}
we see a distinct and subtle difference in the dynamics; gBBM can become negative.  With appropriate initial conditions, an assymetrical global minimum, 
\begin{equation}
\label{eq:comparisonnumericaldata}
u_0=\f_0=1-.9\exp\paren{-\frac{(x-75)^2}{100(1-.999\tanh(x-75))}}
\end{equation}
\eref{eq:bbmcomparison} will cross the axis, while \eref{eq:wellposedmagma} will prop itself up above zero, as pictured Figure \ref{fig:bbmcomparison}.  Away from the minimum in the figures, the solutions are quite similar.  We suspect that it is this nonlinear dispersion that makes \eref{eq:magmaequation} \emph{self-regularizing}.

\begin{figure}
\begin{center}
\label{fig:bbmcomparison}
\caption{The magma and generalized BBM equations, \eref{eq:wellposedmagma} and \eref{eq:bbmcomparison} respectively, evolved from the data, \eref{eq:comparisonnumericaldata}, initially bounded away from zero.  While Magma remains bounded away from zero, BBM not only reaches zero but also becomes negative on a set of positive measure.}
\end{center}
\end{figure}

\subsection{A Comment on Invariant Bounds}
Better understanding the nonlinear dispersive term will likely be necessary to expand the existence result, as we believe the energy estimate approach, on its own, has been exhausted.  Indeed, our estimates, such as \eref{eq:upperboundfunctional} and \eref{eq:lowerboundfunctional} appear quite crude when evaluated numerically and compared to the maximum and minimum values a particular system will actually attain.  

If we start with initial data of the form
\begin{equation}
\f_0(z) = 1+ (1.5-1)\exp{-\frac{(z-100)^2}{2\cdot25}}
\label{eq:postivegaussdata}
\end{equation}
a Gaussian floating on a background of $1$ with a peak at $1.5$, \eref{eq:wellposedmagma} evolves into a solitary wave train, as in \Fref{fig:posgauss}.  The leading solitary wave has an amplitude on the same order as the data, about $1.6$.  
\begin{figure}
        \begin{center}
                \caption{The evolution of data \eref{eq:postivegaussdata} for \eref{eq:wellposedmagma}.  Note that it fails to become significantly larger pointwise from its initial maximum. }
                \label{fig:posgauss}
        \end{center}
\end{figure}
If we were to use \eref{eq:upperboundfunctional} to predict the maximum porosity, we would compute that $\mathcal{U}[\f_0] \sim 10.6$, an order of magnitude higher.

Similarly, \eref{eq:lowerboundfunctional} predicts that the porosity will never go below a value many orders of magnitude smaller than the minimum value the profile eventually evolves into.  For data for the form
\begin{equation}
\f_0(z) = 1- (1-.5)\exp{-\frac{(z-50)^2}{2\cdot25}}
\label{eq:negativegauss}
\end{equation}
corresponding to a negative Gaussian on a background of $1$, with a minimum amplitude of $.5$, it never goes below $.5$.  For the data in \Fref{fig:neggauss}, $\mathcal{L}[\f_0] \sim 8.1e-05$, not to mention that $\mathcal{U}[\f_0] \sim 26$, again in excess of the prediction.
\begin{figure}
        \begin{center}
                \caption{The evolution of data \eref{eq:negativegauss} for \eref{eq:wellposedmagma}.  Note that it fails to become smaller than its initial minimum of $.5$.}
                \label{fig:neggauss}
        \end{center}
\end{figure}

\subsection{Solutions that Reach Zero}
A question that remains is, can we start with smooth data bounded away from zero and reach zero?  While no solution which starts above zero and subsequently reaches zero is known, several functions which are zero on sets of measure zero have been constructed.  So called Compressive Solitary waves, with $\f \leq 1$ and decay to $1$ as $\zeta \to \infty$, are presented in \cite{Nakayama:1991ud, Nakayama:1994cq}.  Some of these are smooth and have a finite number of even roots.  These are certainly classical solutions of \eref{eq:magmaequation}, although the conserved integrals of \Sref{sec:conservation} are not finite for them.  

One function in \cite{Nakayama:1991ud}, along with some of those in \cite{Rabinowicz:2002bj, Takahashi:1992lp} go to zero at cusps, behaving like
\[
\f(z,t)=\f(z-ct) \propto \sqrt{\abs{\zeta}}
\]
near $\zeta = 0$.  Such a function will not have a square integrable derivative, and consequently, as pointed out in \cite{Takahashi:1990tr} for the case $n=3$, $m=0$, the conserved integrals will be also diverge.  Thus these are not solutions in the sense of Definition \ref{def:solution}.

However, they are solutions of the PDE in a weaker sense.  We define a \emph{weak solution} to \eref{eq:magmaequation} as a function $\f(z,t)$ such that
\begin{equation}
\label{eq:weaksolution}
\fl\int_0^\infty \int_{-\infty}^{\infty} \paren{-\dt\psi(z,t) \f(z,t) -\dz\psi(z,t) \f(z,t)^n + \dz \psi(z,t)\f(z,t)^n \dz\paren{ \frac{\dt\f(z,t)}{\f(z,t)^m} }  }dz dt =0
\end{equation}
for all $\psi(z,t) \in C^{\infty}_0(\R \times \R^+)$, and such that both
\begin{eqnarray*}
\f(z,t)\\
\f(z,t)^n \dz\paren{ \frac{\dt\f(z,t)}{\f(z,t)^m} }
\end{eqnarray*}
are in $L^1_{loc}(\R)$ in the $z$ coordinate.  Note that the solutions constructed in \Sref{sec:localexistence} will satisfy these conditions.  

Consider the case $n=3$ and $m=0$, with $c>3$ and define $F$ as follows
\begin{figure}
\label{fig:phaseplot1}
\begin{center}
\caption{Figure (a) shows the phase portait for the traveling wave ODE, \eref{eq:ODE}, of \eref{eq:magmaequation} with $c=4$.  Figures (b) and (c) are weak solutions constructing using parts of the phase portrait.}
\end{center}
\end{figure}

\begin{equation}
\label{eq:cusp}
F(\zeta) = \cases{	\mbox{Curve A of Figure 7 (a)}& for $ \zeta<0$\\
				0& for $ \zeta = 0$\\
				\mbox{Curbe B of Figure 7 (a)}& for $ \zeta>0$}
\end{equation}
Such a function is plotted in Figure 7 (b).  For $\zeta \neq 0$, $F$ solves the second order ODE obtained by making the traveling wave \emph{ansatz} for \eref{eq:magmaequation} and integrating up once with the boundary condition that $F\to 1$ at $\pm\infty$.
\begin{equation}
\label{eq:ODE}
-c F + F^3 +c F^3 F'' = 1-c
\end{equation}
$F$ is also continuous at $\zeta=0$.  Expanding about $\zeta =0$, 
\[
F(\zeta) \sim \sqrt{2}\paren{1-1/c}^{1/4}\sqrt{\abs{\zeta}}
\]
and
\[
\lim_{\abs{\zeta}\to 0, \zeta \neq 0} c F(\zeta)^3 F''(\zeta) = 1-c
\] 
If we define $(c F^3 F'')(0) = 1-c$, then \eref{eq:cusp} satisfies \eref{eq:ODE} as written, \emph{but not}
\[
F'' = \frac{1-c}{c F^3} -\frac{1}{c}+\frac{1}{F^2}
\]

Since $F$ satisifies \eref{eq:ODE} everywhere,
\[
\f(z,t)^3\f(z,t)_{zt} = -c F(z-ct)^3 F''(z-ct) = 1-c -c F(z-ct) + F(z-ct)^3
\]
reducing the left-hand side of \eref{eq:weaksolution} to
\[
\int_0^\infty \int_{-\infty}^{\infty} \paren{-\dt\psi(z,t) -c\dz\psi(z,t) }F(z-ct)dz dt 
\]
This is just the weak form of the transport PDE
\begin{equation}
\label{eq:transport}
\dt u + c \dz u = 0
\end{equation}
which is solved by any function of the form $u(z,t)=u(z- c t)$.  Hence it is a weak solution in the the sense of \eref{eq:weaksolution}.  It is bounded, continuous and in $L^1_{loc}(\R)$.

Unfortunately, there is a uniqueness problem.  Let
\begin{equation}
G(\zeta) = \cases{	\mbox{Curve A of \Fref{fig:phaseplot1}} (a)& for $ \zeta<0$\\
				0 & for $\zeta = 0$\\
				\mbox{Curve C of \Fref{fig:phaseplot1}} (a)& for $ 0<\zeta<z^\star$\\				0 & for $\zeta = z^\star$\\
				\mbox{Curve B of \Fref{fig:phaseplot1}} (a)& for $ \zeta>z^\star$}
\end{equation}
This function is pictured in Figure 7 (c).  Except at $0$ and $z^\star$, $G$ satisifes \eref{eq:ODE} pointwise.  In addition, we claim that, as before, at the two cusps, 
\[
\lim_{\abs{\zeta}\to 0,z^\star, \zeta \neq 0,z^\star} c G(\zeta)^3 F''(\zeta) = 1-c
\] 
so that $G$ in fact satisfies the ODE pointwise, everywhere.  Using the same manipulation as above, we get  
\[
\int_0^\infty \int_{-\infty}^{\infty}  \paren{-\dt\psi(z,t) -c\dz\psi(z,t) }G(z-ct)dz dt
\]
Once again, we have reached the weak form of \eref{eq:transport} and $G$ is a weak solution to this, hence to the magma equation.  Moreover, we could add any number of bumps of finite length of type $C$ between the two curves $A$ and $B$ which go to $1$ at $\pm \infty$.

A selection criterion is needed to distinguish between these functions in the same way an entropy condition is used to distinguish between physical and non-physical solutions to systems of conservation laws.  Although even with such a selection criterion, we recall the physical argument from \cite{Takahashi:1990tr} that there will be stress singularities at a cusps, suggesting none of these are physically realizable.

\ack The authors would like to thank J. Bona, D. McKenzie, and P. Rosenau for their helpful comments.

This work was supported under the National Science Foundation (NSF) Collaboration in Mathematical Geosciences (CMG) grant DMS-0530853, the NSF Integrative Graduate Education and Research Traineeship (IGERT) grant DGE-0221041, and NSF grant DMS-04-12305.


\section*{References}
\bibliography{magma_article}
\end{document}